\documentclass[%
reprint,
 amsmath,amsfonts,amssymb,
 aps,
]{revtex4-2}

\usepackage{graphicx}
\usepackage{dcolumn}
\usepackage{bm}
\usepackage{braket}
\usepackage{siunitx}
\usepackage[english]{babel}

\usepackage{xcolor}

\usepackage{mathtools}

\begin{document}

\preprint{APS/123-QED}
\title{Multipass wide-field phase imager}

\author{\'Alvaro Cuevas}
\email{alvaro.cuevas@icfo.eu}
\affiliation{ICFO-Institut de Ciencies Fotoniques, The Barcelona Institute of Science and Technology, 08860 Castelldefels, Barcelona, Spain}
\author{Daniel Tiemann}
\affiliation{ICFO-Institut de Ciencies Fotoniques, The Barcelona Institute of Science and Technology, 08860 Castelldefels, Barcelona, Spain}
\author{Robin Camphausen}
\affiliation{ICFO-Institut de Ciencies Fotoniques, The Barcelona Institute of Science and Technology, 08860 Castelldefels, Barcelona, Spain}
\author{Iris Cusini}
\affiliation{ICFO-Institut de Ciencies Fotoniques, The Barcelona Institute of Science and Technology, 08860 Castelldefels, Barcelona, Spain}
\affiliation{Politecnico di Milano, Dipartimento di Elettronica, Informazione e Bioingegneria, Piazza Leonardo da Vinci 32, 20133 Milano, Italy}
\author{Antonio Panzani}
\affiliation{Politecnico di Milano, Dipartimento di Elettronica, Informazione e Bioingegneria, Piazza Leonardo da Vinci 32, 20133 Milano, Italy}
\author{Rajdeep Mukherjee}
\affiliation{ICFO-Institut de Ciencies Fotoniques, The Barcelona Institute of Science and Technology, 08860 Castelldefels, Barcelona, Spain}
\affiliation{Karlsruhe Institute of Technology, 76131 Karlsruhe, Germany}
\author{Federica Vila}
\affiliation{Politecnico di Milano, Dipartimento di Elettronica, Informazione e Bioingegneria, Piazza Leonardo da Vinci 32, 20133 Milano, Italy}
\author{Valerio Pruneri}
\affiliation{ICFO-Institut de Ciencies Fotoniques, The Barcelona Institute of Science and Technology, 08860 Castelldefels, Barcelona, Spain}
\affiliation{ICREA-Institució Catalana de Recerca i Estudis Avançats, Passeig Lluís Companys 23, 08010 Barcelona, Spain}

\date{\today}

\begin{abstract}
Advances in optical imaging always look for an increase in sensitivity and resolution among other practicability aspects. Within the same scope, in this work we report a versatile interference contrast imaging technique, capable of sub-\SI{}{\nano\meter} sample-thickness resolution, with a large field-of-view of several $\SI{}{\milli\meter}^{2}$. 
Sensitivity is increased through the use of a self-imaging non-resonant cavity, which causes photons to probe the sample in multiple rounds before being detected, where the configuration can be transmissive or reflective. Phase profiles can be resolved individually for each round thanks to a specially designed single-photon camera with time-of-flight capabilities and true pixels-off gating. Measurement noise is reduced by novel data processing combining the retrieved sample profiles from multiple rounds. Our protocol is specially useful under extremely low light conditions as require by biological or photo-sensitive samples. Results demonstrate at least a five-fold reduction in phase measurement noise, compared to single round imaging, and close values to the predicted sensitivity in case of the best possible cavity configuration, where all photons are maintained until n rounds. We also find a good agreement with the theoretical predictions for low number of rounds, where experimental imperfections would place a minor role. The absence of a laser or cavity lock-in mechanism makes the technique an easy to use inspection tool.
\end{abstract}

\maketitle

\section{Introduction}
In optical metrology and sensing, the search for low noise data is a never ending task, ultimately limited by the Heisenberg uncertainty \cite{Higgins_HeisenbergLimit_2007}. When using linear optics, sensitivity is bounded by the inverse of the information resources available, $\Delta\propto\frac{1}{\sqrt{QM}}$, with $Q$ entangled copies of $M$ number of photons. Without quantum correlations ($Q=1$) we obtain the \textit{shot noise limit} \cite{giovannetti_quantum-enhanced_2004}. Under such classical restriction, noise can only be reduced by increasing $M$ arbitrarily. Unfortunately, this approach requires increasing the power, and represents a high risk for certain chemical bonds that degrade under strong illumination, not just in biological specimens \cite{photo_oxidation1,photo_oxidation2, Popescu_phaseimaging_2011}, but also in synthetic compounds for instance in the optical industry sector \cite{photo_damage}, among others. Alternatively, one can fix the number of photons that probe the sample, while allowing the photons to interact with the sample multiple ($N$) times, effectively providing a sensitivity bound that scales as $\Delta\propto\frac{1}{N}$. In this work we opted for this second approach, while proposing a new protocol that combines the retrieved information from the multiple light-sample interactions, biased on their relative "resource weight". 

Inspection of low optical loss materials was a challenging task until the invention differential interference contrast microscopy (DIM) in the 1960s \cite{lang_nomarski_1982}. With this technique two copies of a coherent light beam are laterally displaced to trespass a target sample at different locations. When recombined later on, beams interfere according to their \textit{optical path difference} (OPD), that can be defined as $OPD=l_{1}n_{1}-l_{2}n_{2}$, with $l_{1}$ ($l_{2}$) and $n_{1}$ ($n_{2}$) as the traveled distance and sample refractive index respectively, at location 1 (2). If both beams have no overlap when trespassing the sample, or the phase is imprinted in one beam only, the phase profile can be retrieved directly by the \textit{phase-shifting digital holography} (PSDH) \cite{gabor_new_1948,malacara_optical_2007}. Assuming an optical interference $I(x,y)=a+b\cos\left(c\phi(x,y)+d\right)$, it requires 4 interference measurements taken with 4 different offsets $\alpha$ to find the parameters $a$, $b$, $c$, $d$, and $\phi(x,y)=\frac{2\pi OPD(x,y)}{\lambda}$. With partial beam overlapping, PSDH does not provide the phase profile directly, but a \textit{sheared phase} $\tilde{\phi}(x,y)$, which requires additional steps to be mapped into $\phi(x,y)$ \cite{Terborg}. During the last decade, DIM has been improved in different aspect, comprising compactness \cite{JUSKAITIS_compactDIC_1994}, larger depth-of-field \cite{haegele_large_2023} and quantum light compatibility \cite{camphausen_2021}. In this work, we go a step forward by combining our multi-pass approach based on a non-resonant cavity that keeps large amount of modes for high transversal resolution together with a no-scanning large field-of-view (FoV) DIM. To the best of our knowledge, multi-pass phase imaging has only been studied so far from the perspective of isolated rounds measurements \cite{juffmann_multi-pass_2016, Israel_multipass_2023}, while multi-pass fluorescence imaging has been studied in combination with active electro-optic control \cite{bowman_electro-optic_2019}.

The requirements of our method are less restrictive than those needed in some modern microscopy techniques. For instance, it can analyse surface roughness or embedded structures in transparent samples, without the need of fluorescent labeling, unlike Stimulated Emission Depletion Microscopy (STED) \cite{Wildanger_STED_2008, Wu_STED_2015}, Photoactivated Localization Microscopy (PALM) or Stochastic Optical Reconstruction Microscopy (STORM) \cite{bonnie_superresolution_2011,Rust_STORM_2006}. Suppose that light dose is limited by the sample photo-sensitivity or the source power up to $M[\SI{}{photons/mm^{2}}]$. By letting pass this amount of light only once through the sample, the phase noise scales like $\Delta\phi\propto\frac{1}{\sqrt{M}}$. If instead, $M/N$ photons interact $N$ times with the sample, the noise scales down to $\Delta\phi\propto\frac{1}{\sqrt{NM}}$ (additional information can be found in \cite{demkowicz-dobrzanski_multi-pass_2010}). Reproducing this condition is especially challenging, because it requires a cavity that closes immediately after light enters, and that opens again at any Nth-round time later. A practical apparatus would require a fast mirror switching, changing from transparent to reflective regime in fractions of a $ns$, following the concept behind \cite{Bowman_switch_2019}, then making it far from a cost effective solution. We instead use a standard permanently open cavity and not only the last accessible round that leaves out of it, but all of them detected sequentially on every acquisition frame. This is possible thanks to a specially designed time-of-flight resolving single-photon avalanche detector (SPAD) array camera, with true pixel-off gating capabilites. The more sensitive technique today involving optical multipass is the cavity ring-down spectroscopy (CRDS) \cite{Sharma_CRDS_2013,Maity_CRDS_2021}. Its working principle is based on quantifying the power decay rate from round to round on the leaking pulses coming out from a cavity. It mainly reveals sample optical absorption and concentration, which can extend to spectral analysis only under resonance conditions, requiring locking mechanisms at laser and cavity levels. In principle it can provide OPD between two modes, but only assisted by a spatially scanning method. We don't reach the CRDS sensitivity levels, but we are able to provide spatial and depth information of the target sample without any scanning mechanism.

\section{Theoretical Model}

Consider a system in which light is prepared in $\ket{H}=(\ket{+}+\ket{-})/\sqrt{2}$ polarization. For every ray, a lateral displacer (LD) splits the components $\ket{+}$ and $\ket{-}$ by a shear distance $d$ and parallel to the input trajectory . The two new rays then enter a non-resonant cavity, which has a partially reflective mirror $M_{R}$ with reflectivity $R$ at the entrance and a fully reflective one ($M$) at the opposite side. By utilizing a cavity round trip $\tau=2L/c$, with length $L$, larger than the light source pulse width ($\Delta t_{p}$) and smaller than the pulse period ($t_{p}$), resonance is prevented. For simplicity reasons, suppose also that the target sample is birefringent, so it adds a phase $\phi$ only in one of the two polarizations. Every time light leaks the cavity, the output polarization can be described by $\ket{\psi_{n}}=(\ket{+}+e^{i\varphi^{(n)}}\ket{-})/\sqrt{2}$ as shown in Fig.~\ref{fig:concept}, with $\varphi_{n}=2(n\phi+\alpha)$, while $\alpha$ is another phase parameter added by the LD. 

\begin{figure}[hbt]
    \centering
    \includegraphics[width=0.45\textwidth]{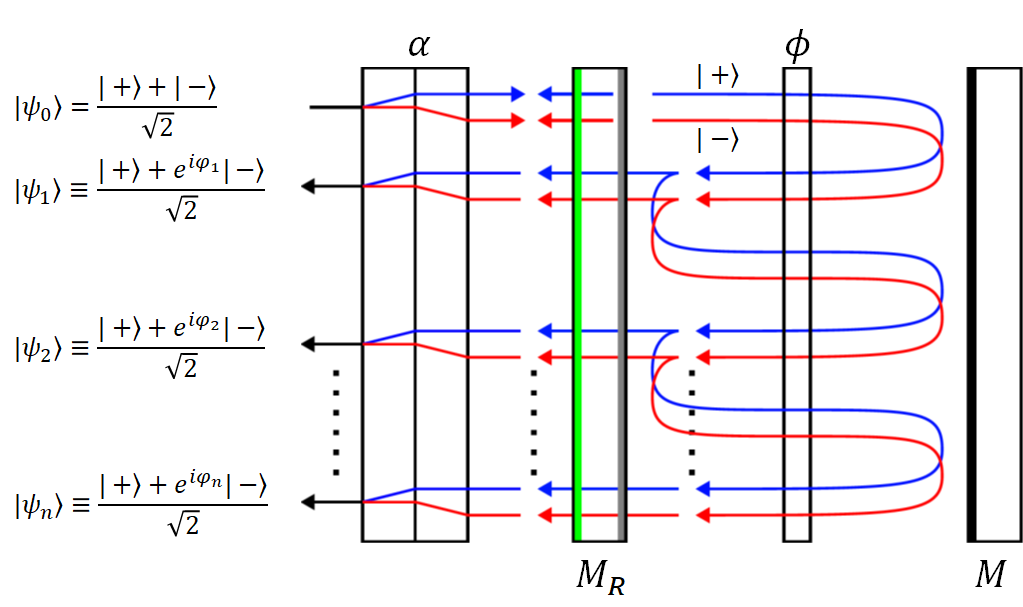}
    \caption{\textbf{Diagram of multi-pass DIM for every light ray}. Input $\ket{H}$-polarized light is split into $\ket{+}$ (blue arrows) and $\ket{-}$ (red arrows) by a birefringent LD before $M_{R}$, and bounces with $M$ inside the cavity. Phases $\phi$ and $\alpha$ are only induced on $\ket{-}$. LD later recombines the outputs.}
    \label{fig:concept}
\end{figure}

Due to the shear effect, an input/output beam in $\mathbf{r}=(x,y)$ coordinate will experience the OPD between $\mathbf{r}'=(x+\delta,y+\delta)$ and $\mathbf{r}''=(x+\delta,y-\delta)$ locations for $\ket{+}$ and $\ket{+}$ respectively, with $\delta=d/\sqrt{2}$. Accordingly, we define the full output state as
\begin{equation}\label{eq:state}
 \ket{\Psi}\approx R^{\frac{1}{2}}\ket{\tau_{0}}\ket{\psi_{0}}+\sum_{n=1}^{N}(1-R)(R\xi)^{\frac{n-1}{2}}\ket{\tau_{n}}\ket{\psi_{n}}
\end{equation}

\noindent with 
\begin{equation}
\ket{\psi_{n}(\mathbf{r})}=(\ket{+(\mathbf{r})}+e^{i2\left(n\tilde{\phi}(\mathbf{r})+\alpha\right)}\ket{-(\mathbf{r})}/\sqrt{2}
\end{equation}

\noindent and $\tilde{\phi}(\mathbf{r})=\phi(\mathbf{r}'')-\phi(\mathbf{r}'))$.

In Eq.\ref{eq:state} we introduce the N-dimensional temporal space that the system can resolve with $\ket{\tau_{n}}=\ket{\tau_{n-1}+\tau}$. The cavity losses are also included in the single-round optical efficiency parameter $\xi$, as well as the double pass factor at every round.

Utilizing PSDH on the system output allows us to retrieve the direct phase estimate
\begin{equation}\label{eq:phase}
    \hat{\theta}^{(n)}=\tan^{-1}\left[\frac{I_{3}^{(n)}-I_{1}^{(n)}}{I_{0}^{(n)}-I_{2}^{(n)}}\right],\\
\end{equation}
and the target super-resolved phase estimate, $\hat{\phi}^{(n)}=\hat{\theta}^{(n)}/2n$ \cite{demkowicz-dobrzanski_multi-pass_2010}. Here the interference amplitudes are obtained by projecting the state into every round-time and the $\ket{V}$ polarization,
\begin{eqnarray}\label{eq:interference}
I_{k}^{(n)}&=&|\bra{\tau_{n}}\braket{V|\Psi}|^{2}\nonumber\\
&=&\frac{\Gamma^{(n)}}{2}\big(1+\mathcal{V}^{(n)}\cos(2\left(n\tilde{\phi}+\alpha_{k})\right)\big).
\end{eqnarray}

This time $\alpha_{k}=\frac{k\pi}{4}$ is a known offset phase imposed by the observer, $\mathcal{V}^{(n)}$ is the interference visibility and $\Gamma^{(n)} =(R\xi)^{n-1}\Gamma^{(1)}= I_{0}^{(n)}+I_{2}^{(n)} = I_{1}^{(n)}+I_{3}^{(n)}$ is the interference amplitude.

Experimentally, each $I_{k}^{(n)}$ corresponds proportionally to the detected photon counts during every acquisitions, $I_{k}^{(n)}=p_{k}^{(n)}$. The uncertainty introduced by the PSDH can then be calculated by computing the standard deviation for the independent $p_{k}^{(n)}$ values through the partial derivatives formula, $\Delta\mathcal{O}=\left(\sum_{k=0}^{3}\sum_{j=1}^{n}\left(\partial \mathcal{O}/\partial p_{k}^{(j)}\cdot\Delta p_{k}^{(j)}\right)^{2}\right)^{1/2}$. From this calculation  (see Supplementary Material for the derivation) we obtain
\begin{equation}\label{eq:noise}
    \Delta\hat{\phi}^{(n)}=\frac{1}{n\mathcal{V}^{(n)}\sqrt{2p^{(n)}}}
\end{equation} 

\noindent where we have introduced the total number of photons detected per round, $p^{(n)}=\sum_{k=0}^{3}p_{k}^{(n)}=2\Gamma^{(n)}$, while using the Poissonian standard deviation $\Delta p_{k}^{(n)}=\sqrt{p_{k}^{(n)}}$.

\begin{figure}[hbt]\label{fig:scaling}
    \centering
    a) \includegraphics[width=0.45\textwidth]{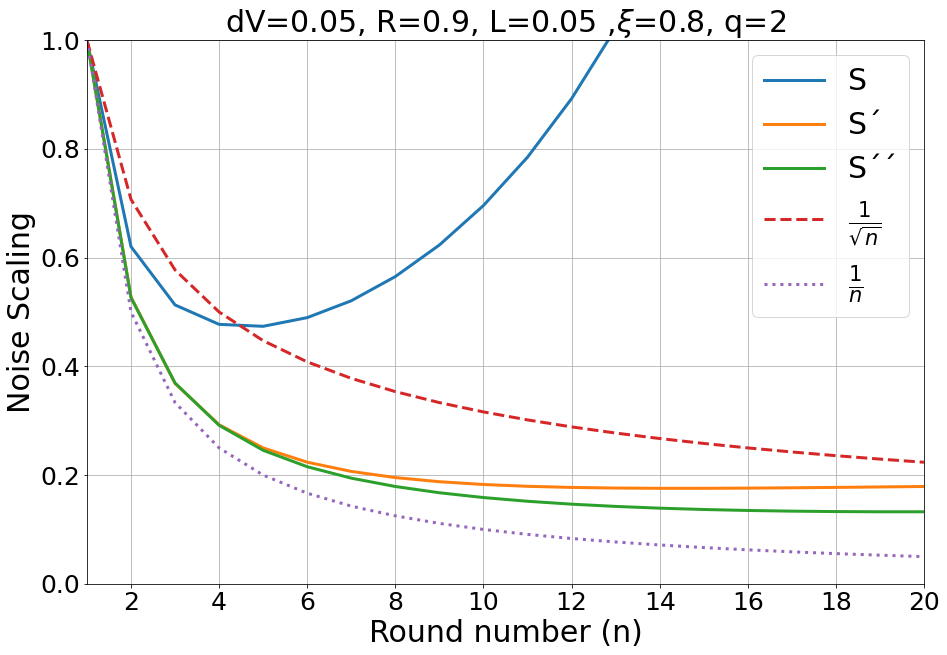}\\
    b) \includegraphics[width=0.45\textwidth]{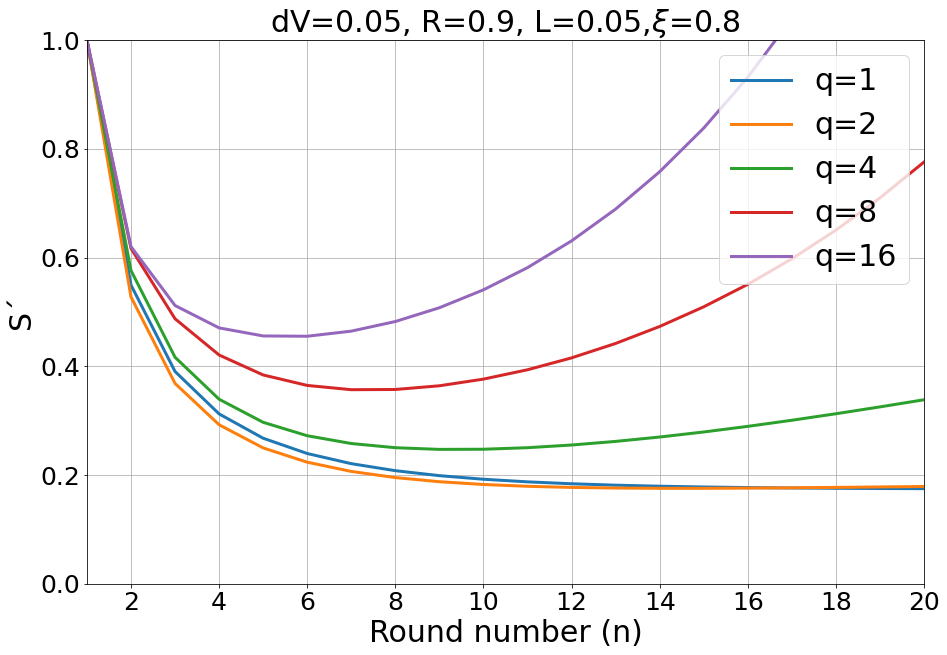}
    \caption{\textbf{Noise scaling simulation}. a) Comparison between PSDH methods. $S'$ outperforms $S$ using realistic optical parameters, while it approaches to $S''$, in which photons don't leak the cavity until round $n$. For the full round trip we considered an efficiency $\xi=(1-L)\xi_{cav}$, being $L=1-R-T$ the partially reflective mirror losses and $\xi_{cav}$ the efficiency of all other optical components within the cavity. We assume an interference visibility that scales like $\mathcal{V}^{(n)}=\left((1-dV)/1\right)^{n-1}$. Curves $1/\sqrt{n}$ and $1/n$ are only used as references, not respect to the shot noise limit. b) Comparison of $S'$ for different values of $q$, where the larger its value, the higher the phase and noise contribution from high rounds number $n$.}
    \label{fig:concept}
\end{figure}

In order to obtain a more abrupt reduction in the phase noise, the next step is to study how phases from different rounds can contribute to it. Accordingly, we propose the following linear combination 
\begin{equation}\label{eq:phase_new}
    \Phi^{(n)}=\sum_{m=1}^{n}w^{(m)}\hat{\phi}^{(m)},\\
\end{equation}

\noindent where weighting factors denote how many resources (interactions with the sample and how many photons) were detected to get each round phase estimate,
\begin{equation}\label{eq:weights}
    w^{(m)}=\frac{m^{q}p^{(m)}}{\sum_{i=1}^{m}i^{q}p^{(i)}}.
\end{equation}

\noindent with $q$ as an arbitrary exponent. The propagated noise from this new method (see Supplementary Material for derivation) will then correspond to 
\begin{equation}\label{eq:noise_new}
    \Delta\Phi^{(n)}=\frac{\sqrt{\sum_{j=1}^{n}\frac{j^{2(q-1)}(R\xi)^{j-1}}{(\mathcal{V}^{(j)})^{2}}}}{\sqrt{2p^{(1)}}\sum_{m=1}^{n}m^{q}(R\xi)^{m-1}}.
\end{equation}

In order to properly evaluate the noise of the multi-pass PSDH, Eq.\ref{eq:noise_new} already includes a fixed number of total photons detected among all 4 phase offset acquisitions. This condition is imposed by the following normalization:
\begin{equation}\label{eq:normalization}
p^{(1)}\longrightarrow p'^{(1)}=p^{(1)}\cdot\frac{p^{(1)}}{\sum_{k=1}^{n}p^{(k)}}=\frac{p^{(1)}}{\sum_{k=1}^{n}(R\xi)^{k-1}}.
\end{equation}

We then define the error scaling of $\hat{\phi}^{(n)}$ and $\Phi^{(n)}$ as:
\begin{eqnarray}\label{eq:scaling1}
    S&=&\frac{\Delta\hat{\phi}^{(n)}}{\Delta\hat{\phi}^{(1)}}=\frac{\mathcal{V}^{(1)}}{n\mathcal{V}^{(n)}(R\xi)^{\frac{n-1}{2}}},\\
    S'&=&\frac{\Delta\Phi^{(n)}}{\Delta\Phi^{(1)}}=\frac{\mathcal{V}^{(1)}}{\sum_{m=1}^{n}m^{q}(R\xi)^{m-1}}\sqrt{\sum_{j=1}^{n}\frac{j^{2(q-1)}(R\xi)^{j-1}}{(\mathcal{V}^{(j)})^{2}}},\nonumber
\end{eqnarray}

\noindent where $\Delta\hat{\phi}^{(1)}=\Delta\Phi^{(1)}=\frac{1}{\mathcal{V}^{(1)}\sqrt{2p^{(1)}}}$.

If we assume all light is preserved until the $n$th round, the exponent in the denominator of $S$ becomes $\frac{n-1}{2}=0$. We define this case as "all light at last round", which is determined by
\begin{equation}\label{eq:scaling2}
    S''=\frac{\mathcal{V}^{(1)}}{n\mathcal{V}^{(n)}},\\ 
\end{equation}

We compare the noise scaling of Eqs.~\ref{eq:scaling1} and Eq.~\ref{eq:scaling2} in Fig.~\ref{fig:scaling}, showing that our method can reduce at least 5 times the phase noise by combining only 5 rounds, and without increasing the number of photons detected.

\section{Experimental Implementation}

\subsection{Optical Apparatus}

We utilized a $\SI{840}{nm}$ Vertical Cavity Surface Emitting Laser (VCSEL) as light probe with a tunable pulse period of $t_{p}>\SI{50}{\mu s}$ and standard deviation width of $\Delta t_p = \SI{200}{ps}$. This source is first coupled into a single-mode fiber (SMF), then collimated with a $1/e^{2}$-diameter of $D=\SI{4}{mm}$ and prepared in $\ket{H}$-polarization by a polarizing beam splitter (PBS). A quarter-wave plate (QWP) and a half-wave plate (HWP) are used before the PBS for controlling the optical power. The LD-operation is implemented by a Savart plate (SP) of thickness $\SI{10}{mm}$, placed with a $\pi/4$-radians rotation to induce a shear of $d=\SI{450}{\mu m}$. The SP is additionally tilted with a motorized stage to further control the $\alpha$ parameter. Transparent samples can be placed immediately after the partially reflective mirror ($M_{R}$) or immediately before the fully reflective mirror ($M$). In case of a reflective sample, $M$ as to be replaced by the sample itself. For a proof of concept, we have chosen $R=0.9$, and a fully variable reflective sample, by utilizing a spatial light modulator (SLM) with $R_{SLM}=0.85$ at $\SI{840}{nm}$ and placed with a $\pi/4$-radians rotation to match the $\ket{+}$ and $\ket{-}$ axes. The cavity optical efficiency is estimated as $\xi_{cav}=R_{SLM}(\xi_{optics})\approx 0.8$, with $R_{SLM}\approx0.88$ and $\xi_{optics}$ including all optical efficiency among lenses and additional components.
\begin{figure}[hbt]
    \centering
    \includegraphics[width=0.48\textwidth]{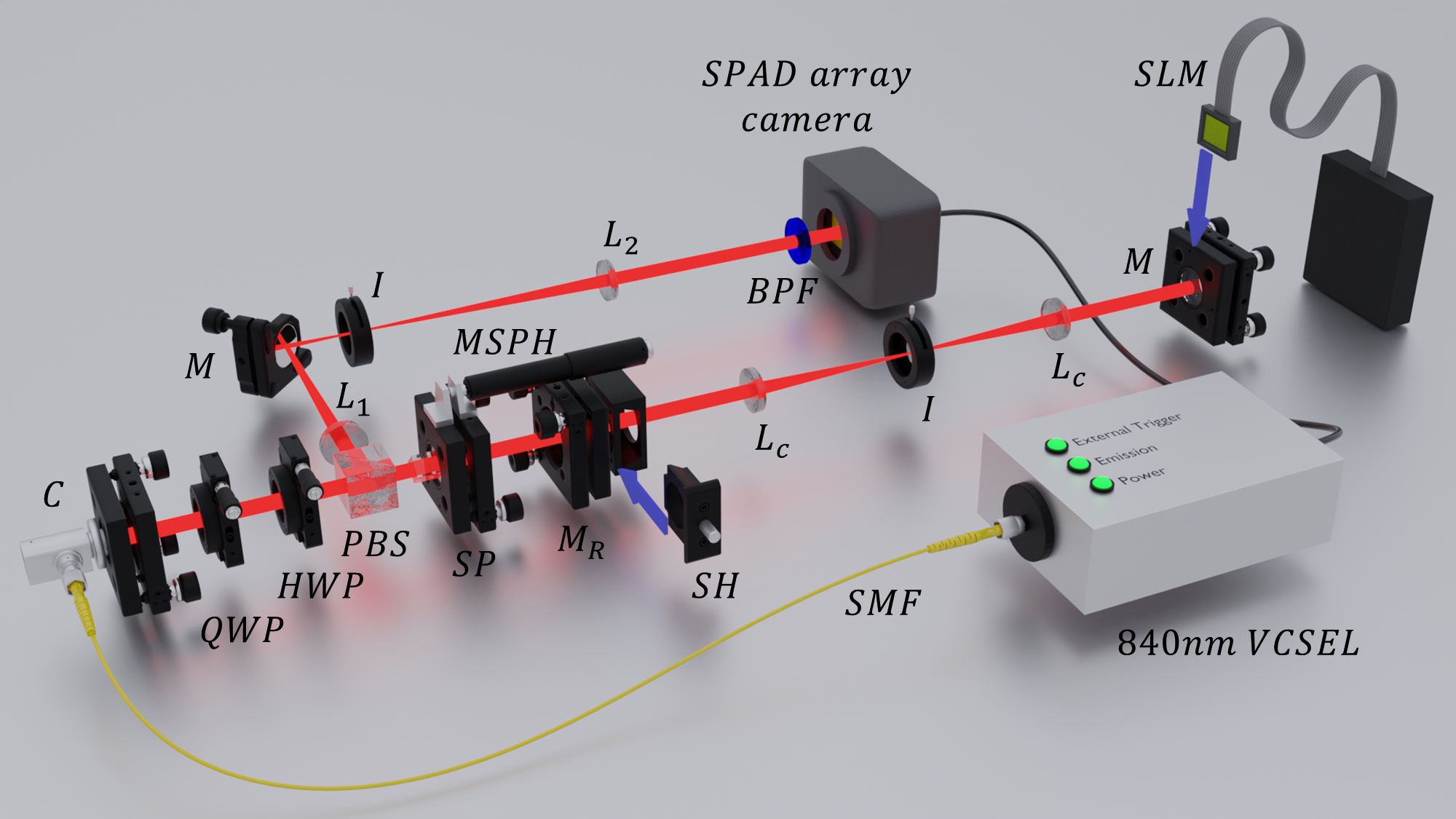}
    \caption{\textbf{Experimental Setup}. Synchronized light pulses are sent from the VCSEL source into the cavity, while the multiple round outputs are registered in the SPAD array camera. Material samples can be inserted through the sample holder (SH), or simulate by the spatial light modulator (SLM). C: collimator; QWP: quarter-wave plate; HWP: half-wave plate, PBS: polarizing beam splitter; $M_{R}$: partially reflective mirror; M: fully reflective mirror; SP: savart plate; MSPH: motorized savart plate holder; I: Iris; $L_{c}$: cavity lenses; $L_{1}$ and $L_{2}$: output telescope lenses; SMF: single-mode fiber; BPF: bandpass filter.}
    \label{fig:setup}
\end{figure}

Figure~\ref{fig:setup} shows the setup, where the cavity has a length $L=\SI{1000}{mm}$, providing the no-resonance condition $\Delta t_{p}<\tau<t_{p}$. Two lenses with focal length $F_{c}=\SI{250}{m}$ are placed in a 4F telescope configuration between $M_{R}$ and $SLM$. The optical interference of Eq. (\ref{eq:interference}) is obtained from the secondary entrance of the system PBS, where output $\ket{V}$ polarization is reflected towards a camera sensor. In this arm, two more lenses with $F_{1}=\SI{200}{mm}$ and $F_{2}=\SI{300}{mm}$ are placed in 4F to image a $\times\frac{3}{2}$ magnified replica of the sample plane into the camera sensor. One pinhole is placed between the two $F_{c}$ lenses and another one in between $F_{1}$ and $F_{2}$ lenses, both for removing scattered light, and $\SI{10}{nm}$ bandpass filter (BPF) is placed in front of the SPAD camera to remove environmental light.

We register the system output with a SPAD array camera, which can recognize every cavity round due to its time-of-flight capabilities. Phase profile estimates are retrieved separately for each round by means of PSDH. Later we combine all phase estimates with the weighting method.  Before performing PSDH, the SP tilting has to be scanned with the cavity optically blocked. Implementing an interpolation similar to Eq. (\ref{eq:interference}) allows to find the motor locations that produce all the $\alpha_{k}$ offsets required by the method.

In order to prevent wrapping effects, we first ensure the direct retrieved phase satisfies the condition $\hat{\theta}^{(n)}<|\pi/2|$. It is in principle easy to exceed this range with the multi-pass approach after many rounds, but knowing $\hat{\theta}^{(n)}=2n\hat{\phi}^{(1)}$ as the expected result, one can implement the experimental correction $\hat{\phi}^{(n)}=\left(\hat{\theta}^{(n)}+\hat{\beta}^{(n)}\right)/2n$, with
\begin{equation}
\hat{\beta}^{(n)}=\Bigg\{\begin{matrix}
\frac{-\pi}{n}&\text{if}& \hat{\theta}^{(n)}-2n\hat{\phi}^{(1)}>\pi\\
0&\text{if}&|\hat{\theta}^{(n)}-2n\hat{\phi}^{(1)}|\leq\pi\\
\frac{\pi}{n}&\text{if}& \hat{\theta}^{(n)}-2n\hat{\phi}^{(1)}<-\pi
\end{matrix}.
\end{equation}

\subsection{Image Sensor}
The SPAD camera we have employed is based on a revised version of the SPAD chip presented in \cite{Portaluppi}, which is designed for single-photon timing and counting. The chip has been manufactured in a 0.16 $\mu$m BCD (Bipolar-CMOS-DMOS) SPAD technology and integrates an array of 32 x 32 square SPADs (32 $\mu$m × 32 $\mu$m) with 100 $\mu$m pitch, resulting in a 9.6\% fill factor. The photon-detection probability is about 10\% at 840 nm and the average dark count rate is 700 cps at 5 V excess bias and room temperature. 

The front-end circuit is a gated variable load quenching circuit designed to quickly turn on/off the detectors \cite{tisa2007electronics}. With this capability we are able, while triggering the VCSEL with it, to keep the pixels off during a tunable time to prevent detection of the pulse immediately reflected by $M_{R}$ before entering the cavity. This signal effectively arrives to the SPADs array and could blind it, preventing us to measure large number of rounds. We have experimentally verified that the average duration of the gate rising edge (10-90\% transition) is 340 ps. 

256 time-to-digital converters, with 204.8 $\mu$s full-scale range and 50 ps resolution, are shared among the pixels and a discriminator circuit is employed to preserve the spatial resolution. The complete camera (9×7×5 cm$^3$) is based on an FPGA and transfers the data to the PC through a USB 3.0 connector allowing frame rates up to 100 kfps. The gate window position can be set with 80 ps precision, in this way the unwanted reflection from the first mirror can be completely filtered out without hiding the first useful peak.

\subsection{Acquisition and data conversion}

The raw data from the SPADs array camera is first stored in HDF5 format, with frame index, timestamp and pixel index of each photon count. A PSDH dataset consists of two batches, each containing four $\alpha$-offset acquisitions, one batch recorded with the sample and the other without it, so we can determine background information too. We define fixed time intervals around each peak to count the effective round photon detections $p_k^{(n)}(x,y)$. The number of frames $f$ throughout the eight independent acquisitions of the PSDH dataset may not be identical. Therefore, explicitly only the event counts up to the largest common frame count $f_{max}$ available across all acquisitions in the dataset are used. From each dataset batch, the sum in photon counts across all four $\alpha$-offsets, $p^{(n)}(x,y)$, is extracted for each round $n$. In order to satisfy the normalization condition imposed by Eq. (\ref{eq:normalization}), in which photon counts (of all pixels combined) from many rounds equal the ones detected in the first round $p^{(1)}$, data subsets of the overall acquisition have to be prepared. This is done by reducing the number of used frames instead of photon counts directly, preventing a computationally biased filtering. In a first step, the such needed frames are estimated by
$f\rightarrow f^{(n)} \approx f$\ldots$ p^{(1)}/\sum_{m=1}^{n} p^{(m)}$ and subsequently the result is optimised by adjusting $f^{(n)}$ iteratively for the best match to the target photon count $p^{(1)}$. For each number of rounds $n$ in the dataset, the required frame count $f^{(n)}$ is determined; and with them, the photon counts per pixel are extracted from the dataset for each round $m$ ($m = 1,\ldots,n$), which are considered as the normalised photon counts $p^{\prime(m)}(x,y)$.

\section{Results}

When measuring non-birefringent samples either in transmission or reflection mode, the method retrieves a sheared phase profile. Alternatively, for birefringent samples, the method retrieves the "true" phase profile. In our proof of principle we opted for this second case, where the SLM imprints phase only in one of the two polarized beams within the cavity. Accordingly, we can fairly assume the absence of shear in the process ($\tilde{\phi}^{(n)}\equiv \phi^{(n)}$). 

\begin{figure}[hbt]
    \centering
    \includegraphics[width=0.48\textwidth]{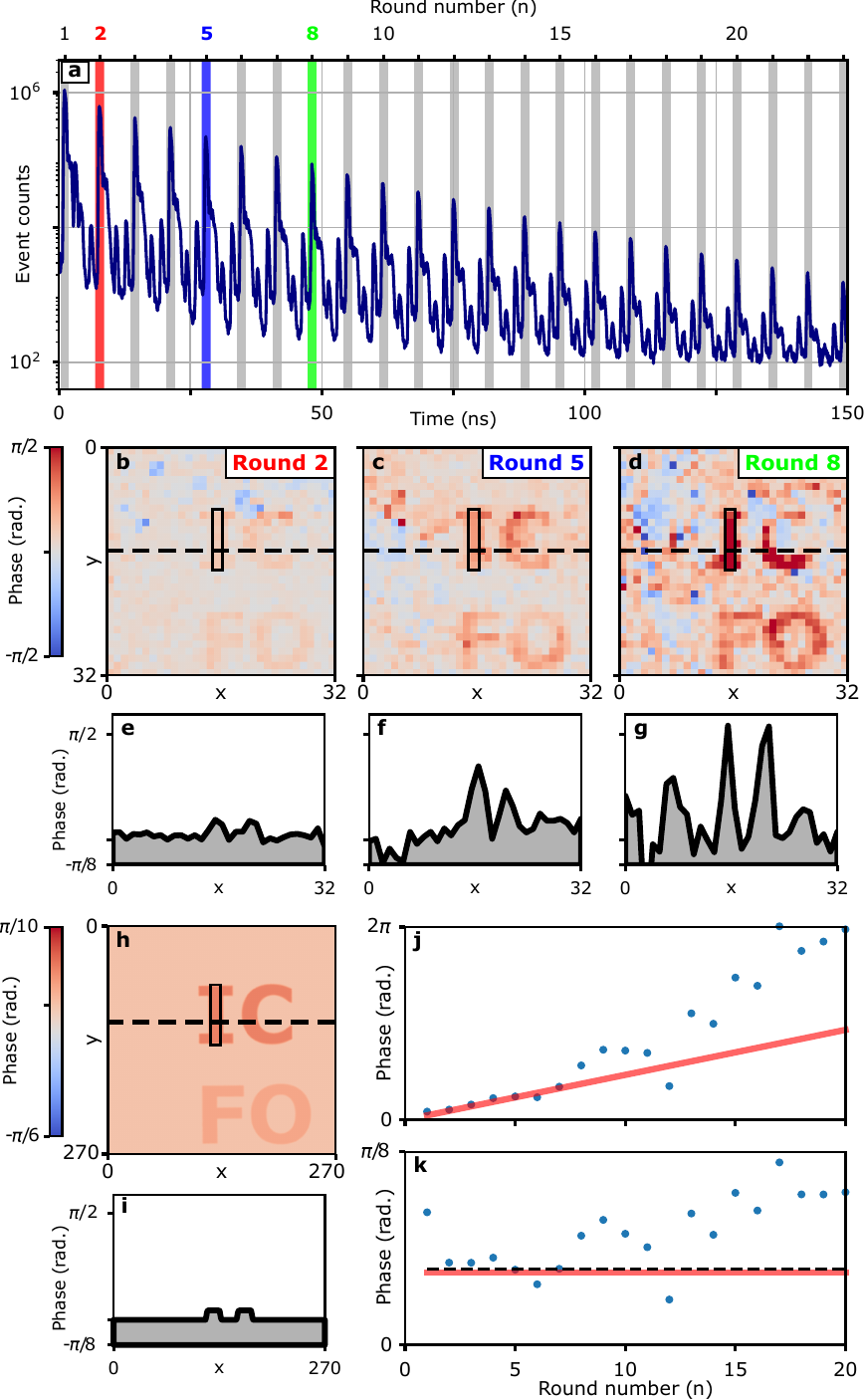}
    \caption{\textbf{System calibration results}. a) Temporal slicing of the detected pulse train. Direct phase profiles $\hat{\theta}^{(n)}$ for 2nd b), 5th c) and 8th d). e), f) and g) are cross-sections of b), c) and d), respectively. h) is the projected ideal phase profile, showing features with depth $\phi\approx\pi/42$ and $\phi\approx\pi/21$ over a background of $\phi=0$, together with its cross-section in i). j) Average values of $n\phi^{(n)}$ for the rectangle regions of interest around the letter "I" with a linear prediction in red. k) Average values of $\phi^{(n)}$ for the same regions of interest, with a linear prediction in red.}
    \label{fig:calibration_results}
\end{figure}

\begin{figure}[hbt]
    \centering
    \includegraphics[width=0.48\textwidth]{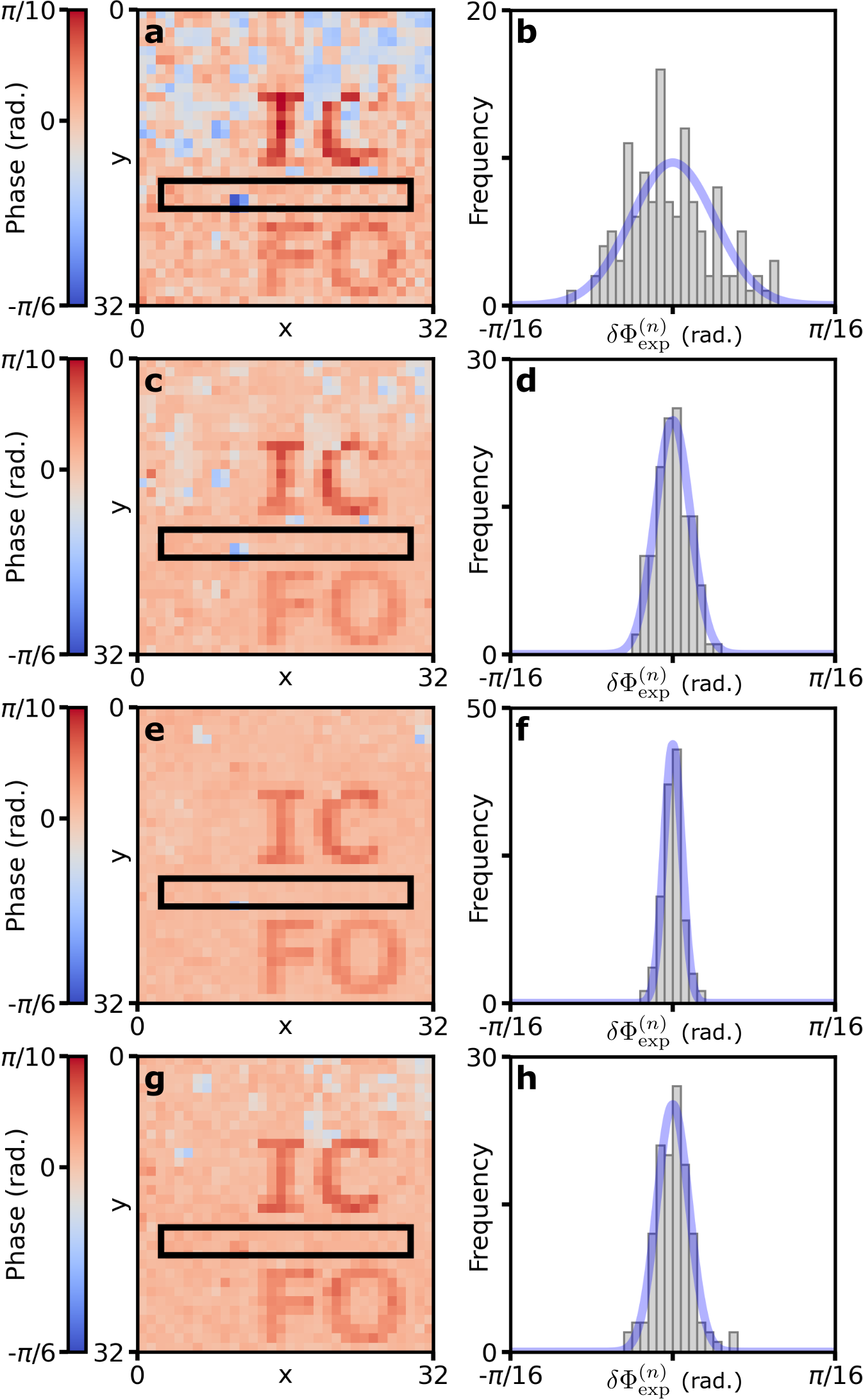}
    \caption{\textbf{Multi-pass phase profiles results.} Left column: Experimental phase images $\Phi^{(n)}$ retrieved by combining $n$ individual rounds (Eq.~(\ref{eq:phase_new}) and Eq. (\ref{eq:weights}), with $q=2$). Right column: Histograms of pixel-to-pixel noise in retrieved phase images $\delta\Phi^{(n)}$, i.e., differences between neighbouring pixels in black rectangles in left column; solid blue line: Gaussian fitting. a), b) $n=1$. c), d) $n=3$. e), f) $n=7$. g), h) $n=17$.}

    \label{fig:multipass_round_profiles}
\end{figure}
In Fig.~\ref{fig:calibration_results}a) we show a typical histogram of accumulated photon counts, where we implement a $\SI{1}{ns}$ gating around each peak. Working at high time-resolution ($\SI{50}{ps}$ per bin) allows us to better remove side peaks which might contain spurious phase profiles, generated for instance from the lenses' back-reflection or even from outer cavity components. In Figs.~\ref{fig:calibration_results}b), c) and d) we show the direct retrieved phase profiles $\hat{\theta}^{(n)}$ for the 2nd, 5th and 8th rounds, respectively. In Figs.~\ref{fig:calibration_results}e), f) and g) we show cross sections of the above profiles, together with the ideal SLM projection and its cross section in h) and i), respectively. To better illustrate the increasing values of this retrieved phase we selected a rectangular region of interest (ROI) around the "I" letter in our sample, and from here we extracted the average phase values, which we plot as $\hat{\theta}^{(n)}$ in Fig.~\ref{fig:calibration_results}j) and $\hat{\phi}^{(n)}$ in Fig.~\ref{fig:calibration_results}k). Here we see that starting at the 8th round the retrieved phases deviate from the predicted values. 

Figure~\ref{fig:multipass_round_profiles} shows how phase measurement sensitivity is enhanced by our method. a), c), e) and g) show the experimentally retrieved phase profiles $\Phi^{(n)}$ (calculated from Eq.~(\ref{eq:phase_new})) for $n=1$, $n=3$, $n=7$ and $n=17$ rounds combined, respectively. b), d), f) and h) show the histogram of pixel-to-pixel differences, $\delta\Phi^{(n)}$, for the rectangular ROI in a), c), e) and g), respectively. This result was obtained with an equal number of total photon detections in all cases. On the left column of Fig.~\ref{fig:multipass_round_profiles} we see that $\Phi^{7}$ present the smoother profile, because it contains the maximum number of rounds in which $\phi^{(n)}$ does not deviate from the linear prediction, as shown above in Figs.~\ref{fig:calibration_results}j) and k). On the right column of Fig.~\ref{fig:multipass_round_profiles} we verify the above since $\delta\Phi^{(7)}$ presents the narrower values. The standard deviation parameter obtained by fitting a Gaussian curve to each $\delta\Phi^{(n)}$ histogram (transparent light blue curves in Fig.~\ref{fig:multipass_round_profiles}) yields a quantitative measure of $\Delta\Phi^{(n)}$; this empirical measure of image noise is also known as the local uncertainty \cite{camphausen_2021, Israel_NOON_2014}. These results were obtained by an absolute number of photons detected equal to $p=1.109\cdot 10^{7}$ among all the four phase-offsets for sample plus background. This was done by accumulating over $f=62353$ frames, each one with length $t_f=\SI{500}{\mu s}$, giving us a total acquisition time of $\SI{31.2}{s}$. The four phase-offset acquisitions for background only would require a similar amount of time to collect the same number of photons. The arbitrary photon count rate, to stay under saturation regime, was $3.55\cdot 10^{5} \textnormal{photons}/s$.

\begin{figure}[hbt]
    \centering
    \includegraphics[width=0.48\textwidth]{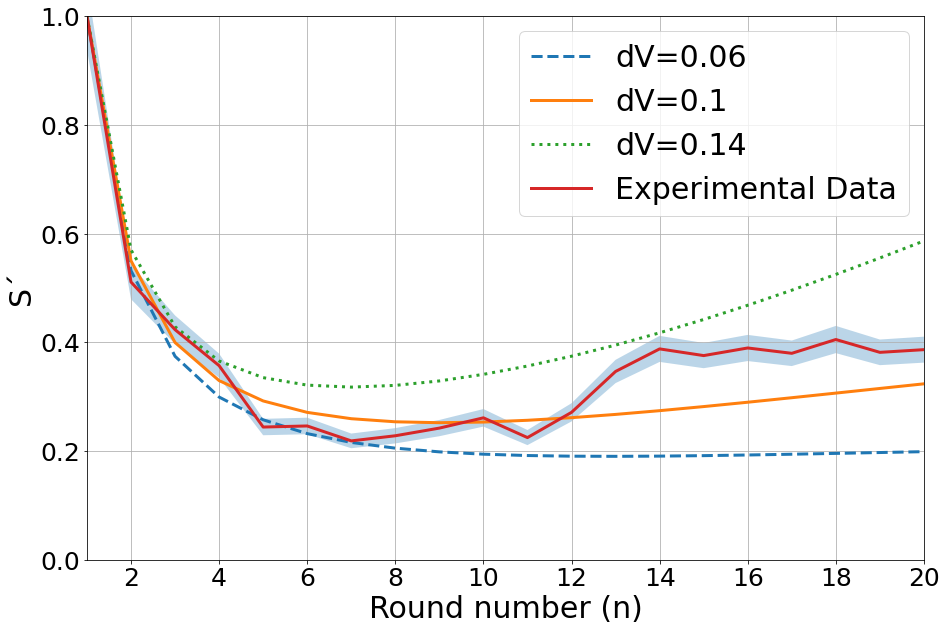}
    \caption{\textbf{Phase noise reduction enabled by multipass PSDH}. Results evaluated by LU from the selected region of interest shown in Fig.~\ref{fig:multipass_round_profiles}. Simulated curves as reference using experimental parameters as $R=0.05$, $L=0.05$, $\xi=0.8$, with $q=2$ and different values for the visibility decay. }
    \label{fig:local_uncertainty}
\end{figure}

The reduction in experimental phase measurement uncertainty, enabled by our multi-pass combination method, can be quantified by comparing with the single-pass noise $\Delta\Phi^{(1)}$ (i.e., standard deviation of Gaussian fitting function in Fig.~\ref{fig:multipass_round_profiles}b)).
Figure~\ref{fig:local_uncertainty} plots the normalized measurement noise $S'(n) \equiv \Delta\Phi^{(n)}/\Delta\Phi^{(1)}$ as a function of $n$, the number of rounds combined (solid blue line). The grey shaded area represents the statistical standard error in $S'(n)$. As seen in Fig.~\ref{fig:local_uncertainty}, our method strongly reduces phase measurement uncertainty, approaching closely to the ideal cavity case lower bounded by $1/n$. $\Delta\Phi^{(n)}$ reaches a minimum for $n=7$, yielding a measurement uncertainty reduction by a factor of $0.22 \pm 0.01$ compared to the single-pass case $\Delta\Phi^{(1)}$.
We also observe in Fig.~\ref{fig:local_uncertainty} that for $n\gtrsim 10$ the normalized measurement noise $S'$ increases again. This is attributed to a loss of interference visibility as round number increases, due to imperfect optical alignment causing the probe beam to drift laterally over many rounds. The experimental noise reduction curve $S'(n)$ is consistent with our theoretical model.

\section{Conclusion}

Our multi-pass approach combined with PSDH imaging is able to reduce the measurement noise down to 0.22 when compared to the single-pass noise in less than 10 rounds. Theoretically, we can improve this value far beyond, as shown in the Supplementary Material, by modifying system parameters like the cavity efficiency, mirror reflectivity and losses, etc. We compared the noise scaling versus other PSDH alternatives, showing how resilient and close to the ideal scenario of no photon decay can be. In order to validate this comparison, the number of photons detected was fixed to a constant number for all experiments. This restriction was implemented computationally by only removing frames from the full dataset, while applying no bias protocols. 

We believe the discrepancy between our simulations and the experimental results after round 7 could be explained by a series of factors; i) an increasing influence of the array hot pixels, which have not been deactivated during our tests; ii) the always present dark counts rate and potential cross-talk, which has not been simulated in our model; iii) the Gaussian beam expansion in a $\SI{2}{\meter}$ full round trip cavity, that might prevent photons to always pass through the sample in the same transverse position; iv) the difficulties in the mirrors alignment in absence of transverse and longitudinal modes interference. We believe that i) and ii) cannot be fully overcome by increasing pump power. This choice would produce pixel saturation in very few rounds, then limiting the applicability of the method. With the current components, the cavity size could be reduced by one order of magnitude to make the system more portable and then reduce influence of the beam expansion. Further miniaturization would require a much lower jitter imager, a technology which is not yet available. Lastly, iv) can be overcome by programming the imager FPGA to get live view of the different cavity rounds separately. We consider to explore these improvements in future experimental iterations as well as to theoretical prediction of the sensitivity lower bounds, considering all the above additional restrictions.

One major advantage of the protocol is that it allows large field of view imaging compatible with sample and/or camera scanning, which can increase the (x,y)-resolution as much as the user desires. Additionally, the system can also be used for monitoring fast optical effects (with ns-scale variations) inside the cavity, where our gating mechanism can provide temporal slices of the studied phenomenon. In such a scenario, an nth-gate  will have the accumulated information from all previous gates. One can iteratively isolate the nth-gate information only to finally obtain video recordings, as long as the observed phenomenon is optically/electronically triggered in synchronicity with the camera. 

Overall, we provide a unique combination of technologies for super-sensitive phase imaging, designed as an optical inspection tool compatible with reflective and transparent samples with birefringent and/or non-birefringent features. Quantum imaging alternatives, that utilize either multi-photon entanglement or squeezed states of light \cite{casacio_quantum-enhanced_2021}, can  provide sensitivities on the same order of magnitude or higher. However, at present such quantum-enhanced approaches are impractical to scale to real-world advantages, due to demanding experimental requirements such as very high optical efficiencies. \\

\newpage
\section{Supplementary Material}
\subsection{Derivation of Noise Decay}
In order to calculate the error of the multi-pass phase $\Phi^{(n)}$ propagated from photon count fluctuations one has to first calculate the error of any round retrieved phase $\hat{\phi}^{(n)}$. Let's start with the relation between the photon counts $p_{j}^{(n)}$ and the phase offsets $\alpha_{k}=\frac{k\pi}{4}$:

\begin{widetext}
\begin{eqnarray}
p_{0}^{(n)} &=& \frac{p^{(n)}}{4}\left(1+\mathcal{V}^{(n)}\cos\left(2\left(n\phi^{(n)}+\alpha_{0}\right)\right)\right) = \frac{p^{(n)}}{4} (1+\mathcal{V}^{(n)}\cos(2n\phi^{(n)})), \nonumber\\
p_{1}^{(n)} &=& \frac{p^{(n)}}{4}\left(1+\mathcal{V}^{(n)}\cos\left(2\left(n\phi^{(n)}+\alpha_{1}\right)\right)\right) = \frac{p^{(n)}}{4} (1-\mathcal{V}^{(n)}\sin(2n\phi^{(n)})), \nonumber\\
p_{2}^{(n)} &=& \frac{p^{(n)}}{4}\left(1+\mathcal{V}^{(n)}\cos\left(2\left(n\phi^{(n)}+\alpha_{2}\right)\right)\right) = \frac{p^{(n)}}{4} (1-\mathcal{V}^{(n)}\cos(2n\phi^{(n)})), \nonumber\\
p_{3}^{(n)} &=& \frac{p^{(n)}}{4}\left(1+\mathcal{V}^{(n)}\cos\left(2\left(n\phi^{(n)}+\alpha_{3}\right)\right)\right) = \frac{p^{(n)}}{4} (1+\mathcal{V}^{(n)}\sin(2n\phi^{(n)})),
\end{eqnarray}
\end{widetext}
where $\mathcal{V}^{(n)}$ is the interference visibility at every round. Since the proof of principle is implemented by an SLM device acting only in one polarization, we can fairly assume that there is no shear in the process ($\tilde{\phi}^{(n)}\equiv \phi^{(n)}$).

Accordingly, the partial derivatives on $\hat{\phi}^{(n)}$ can be written as

\begin{eqnarray}
\frac{\partial \hat{\phi}^{(n)}}{\partial p_{0}^{(n)}}&=&\frac{-(p_{3}^{(n)}-p_{1}^{(n)})}{\left(p_{0}^{(n)}-p_{2}^{(n)}\right)^{2}+\left(p_{3}^{(n)}-p_{1}^{(n)}\right)^{2}}= \frac{\sin}{n\mathcal{V}^{(n)}p^{(n)}},\nonumber\\
\frac{\partial \hat{\phi}^{(n)}}{\partial p_{1}^{(n)}}&=&\frac{-(p_{2}^{(n)}-p_{0}^{(n)})}{\left(p_{0}^{(n)}-p_{2}^{(n)}\right)^{2}+\left(p_{3}^{(n)}-p_{1}^{(n)}\right)^{2}}= \frac{-\cos}{n\mathcal{V}^{(n)}p^{(n)}},\nonumber\\
\frac{\partial \hat{\phi}^{(n)}}{\partial p_{2}^{(n)}}&=&-\frac{\partial \hat{\phi}^{(n)}}{\partial p_{0}^{(n)}}= \frac{\sin}{n\mathcal{V}^{(n)}p^{(n)}},\nonumber\\
\frac{\partial \hat{\phi}^{(n)}}{\partial p_{3}^{(n)}}&=&-\frac{\partial \hat{\phi}^{(n)}}{\partial p_{1}^{(n)}}= \frac{\cos}{n\mathcal{V}^{(n)}p^{(n)}},
\end{eqnarray}

The phase error for any round will then be defined as
\begin{widetext}
\begin{eqnarray}
    \Delta\hat{\phi}^{(n)}&=&\sqrt{\sum_{j=0}^{3}\left(\frac{\partial \hat{\phi}^{(n)}}{\partial p_{j}^{(n)}}\right)^{2}\cdot\left(\Delta p_{j}^{(n)}\right)^{2}}\nonumber\\
    &=&\frac{1}{2n\mathcal{V}^{(n)}p^{(n)}}\Big[\sin^{2}(2n\phi^{(n)})p^{(n)}\left((1+\mathcal{V}^{(n)}\cos(2n\phi^{(n)}))+(1-\mathcal{V}^{(n)}\cos(2n\phi^{(n)}))\right)\nonumber\\
    && +\cos^{2}(2n\phi^{(n)})p^{(n)}\left((1-\mathcal{V}^{(n)}\sin(2n\phi^{(n)}))+(1+\mathcal{V}^{(n)}\sin(2n\phi^{(n)}))\right)\Big]^{1/2}\nonumber\\
    &=&\frac{1}{n\mathcal{V}^{(n)}\sqrt{2p^{(n)}}}
\end{eqnarray} 
\end{widetext}

where we have considered the photon counts error as the Poissonian standard deviation, $\Delta p_{j}^{(n)} = \sqrt{p_{j}^{(n)}}$.

If we assume the weight of each round $\omega^{(j)}$ can be predicted in advance given the known optical decay for every round, $p^{(n)}=p^{(1)}(R\xi)^{n-1}$, it is valid to consider $\Delta\omega^{(j)}=0$. Then, the multi-pass phase error will be defined as
\begin{widetext}
\begin{eqnarray}
    \Delta \Phi^{(n)} &=& \sqrt{\sum_{j=1}^{n}\left(\frac{\partial \Phi^{(n)}}{\partial \hat{\phi}^{(j)}}\right)^{2}\cdot \left(\Delta \hat{\phi}^{(j)}\right)^{2}}\nonumber\\
    &=&\sqrt{\frac{1}{(1\cdot\mathcal{V}^{(1)})^{2}2p^{(1)}}\cdot\left(\frac{1^{q}\cdot p^{(1)}}{\sum_{m=1}^{n}m^{q}p^{(m)}}\right)^{2}+\cdots+\frac{1}{(n\cdot\mathcal{V}^{(n)})^{2}2p^{(n)}}\cdot\left(\frac{n^{q}\cdot p^{(n)}}{\sum_{m=1}^{n}m^{q}p^{(m)}}\right)^{2}}\nonumber\\
    &=&\frac{1}{\sqrt{2}\sum_{m=1}^{n}m^{q}p^{(m)}}\sqrt{\frac{1^{2(q-1)}p^{(1)}}{(\mathcal{V}^{(1)})^{2}}+\cdots+\frac{n^{2(q-1)}p^{(n)}}{(\mathcal{V}^{(n)})^{2}}}\nonumber\\
    &=&\frac{1}{\sqrt{2p^{(1)}}\sum_{m=1}^{n}m^{q}(R\xi)^{m-1}}\sqrt{\sum_{j=1}^{n}\frac{j^{2(q-1)}(R\xi)^{j-1}}{(\mathcal{V}^{(j)})^{2}}},
\end{eqnarray}
\end{widetext}

where $q$ is an arbitrary modulation factor, $R<1$ is the cavity reflectivity and $\xi$ is the cavity optical efficiency.

\begin{figure}[bht]
    \centering
    \includegraphics[width=0.48\textwidth]{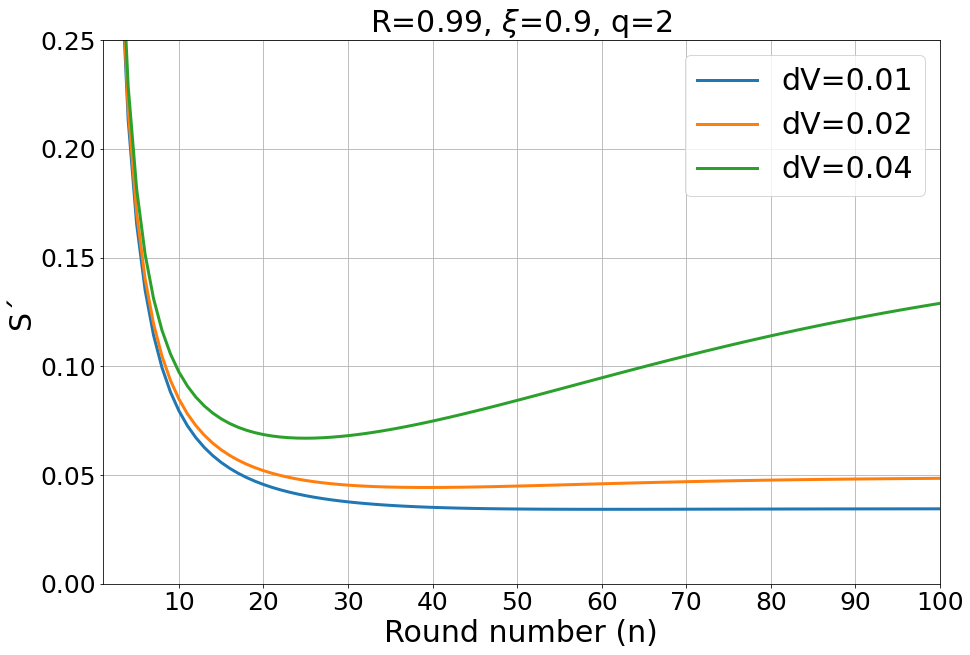}
    \caption{\textbf{Achievable noise reduction of the method under optimal experimental conditions.} We compare the noise decay under three scenarios of interference degradation, with $1\%$, $2\%$ and $4\%$ visibility reduction at every round respect to the previous one.}
    \label{fig:supplementary1}
\end{figure}

In Fig.~\ref{fig:supplementary1} we show a simulation of the expected noise decay for our method, $S'=\frac{\Delta \Phi^{(n)}}{\Delta \Phi^{(1)}}=\frac{\Delta \Phi^{(n)}}{\Delta \hat{\phi}^{(1)}}$, considering optimized experimental scenarios in which the mirror reflectivity is R=0.99 and the overall optical efficiency is $\xi=0.9$, all optical components considered including sample transmission. In the best case we can reach noise levels under $5\%$ in less than 20 rounds when compared to single-pass phase imaging.

\subsection{SPADs saturation and Hard-gating.}

\begin{figure}[hbt]
    \centering
    \includegraphics[width=0.48\textwidth]{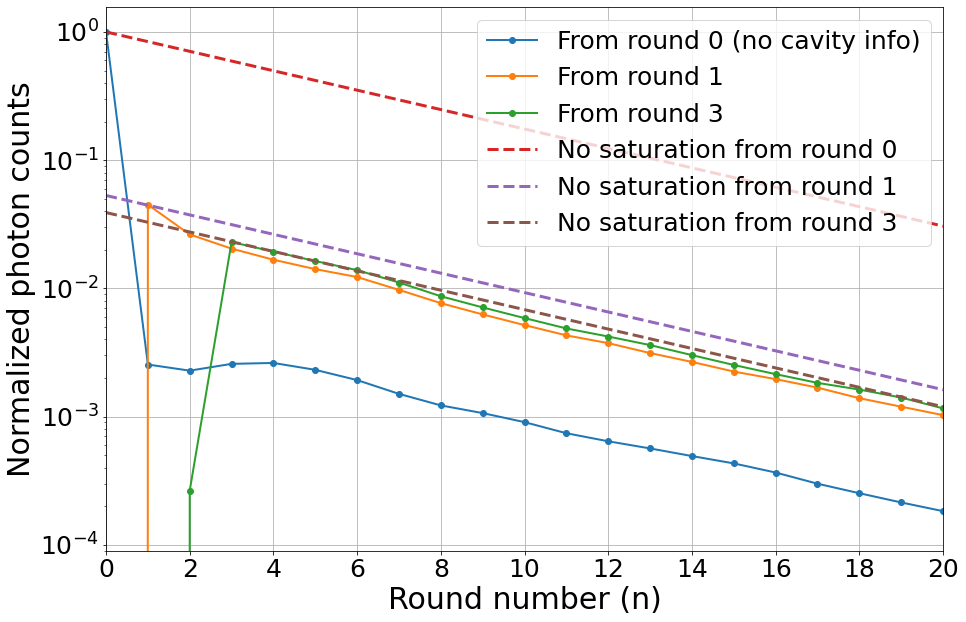}
    \caption{\textbf{Pulse decay detection for different starting rounds normalized to maximum photon counts}. Connected dot-line curves represent standard experimental data. Segmented lines represent simulated photon decay if there is no saturation and using $\eta=0.84$.}
    \label{fig:supplementary}
\end{figure}

In Fig.~\ref{fig:supplementary} we show the effects of activating the camera SPADs at different times, detecting light from the unuseful pulse that do not enter the cavity, the first one that leaves the cavity and the second one that leaves the cavity. The data correspond to photon counts integrated over more than 1000 frames across the entire SPAD array within $\SI{1}{ns}$ gates and normalized to the maximum number of counts. A no saturation condition should coincide with an exponential decay (here depicted as a line in log scale), $counts=A\cdot \eta^{n}$, with $A$ as an arbitrary amplitude and $\eta$ as the effective system optical efficiency. Instead, we see that activating pixels from the unused pulse the photon counts plot does not follow such a decay. It start with high number of counts, it continues with a low number of counts and it finally decays exponentially. Activating pixels from round 1 it approaches closely to the exponential decay, which is already fulfilled when pixels are activated from round 2.

Preventing saturation by hard gating allows to properly implement the weighted phase combination, which depends on the relative number of photons among rounds, and to also approach to the number of photons that actually interact with the samples. Without this camera feature, the multi-pass sensitivity enhancement would never work.

\newpage
\textbf{Funding:} Agència de Gestió d'Ajuts Universitaris i de Recerca (2021 SGR 01458); H2020 Marie Skłodowska-Curie Actions (713729 (ICFOstepstone 2)); H2020 Future and Emerging Technologies (801060 (Q-MIC)); Generalitat de Catalunya; Centres de Recerca de Catalunya; FUNDACIÓ Privada MIR-PUIG; Fundación Cellex; Agencia Estatal de Investigación (CEX2019-000910-S, CEX2019-000910-S (MCIN/ AEI/10.13039/501100011033), PID2019-106892RB-I00); European Union Next Generation (PRTR-C17.I1).\\

\textbf{Acknowledgements:} This study was supported by MICIIN with funding from European Union NextGenerationEU (PRTR-C17.I1) and by Generalitat de Catalunya.\\

\textbf{Author Contributions:} \'A.C. proposed the experiment and developed the theoretical model; D.T., R.C and R.M. performed the experiment and developed the data processing code; I.C. programmed the SPAD array FPGA and implemented critical integration aspects of the experiment; A.P. and F.V. developed the SPAD camera; V.P. supervised the experiment and technological perspectives. All authors participated in the manuscript writing.\\

\textbf{Competing interests:} 
No competing interests.\\

\textbf{Data availability:} Data underlying the results presented in this paper are available from the corresponding author upon reasonable request.


%

\end{document}